\documentclass[runningheads]{llncs}
\usepackage{graphicx}
\usepackage{enumitem}
\setlist{nosep, leftmargin=14pt}

\usepackage{mwe} 
\usepackage{tabularx,booktabs,tikz}
\usepackage{color}
\usepackage{amsmath,amssymb,amsfonts}
\usepackage{siunitx}
\usepackage{hyperref}
\begin{document}
\title{Efficient Model Monitoring for Quality Control in Cardiac Image Segmentation\thanks{This work was supported through funding from the Monaco Government.}}
\titlerunning{Quality Control in Cardiac Image Segmentation}
%
\author{Francesco Galati\inst{1} \and
Maria A. Zuluaga\inst{1}\orcidID{0000-0002-1147-766X}}
\authorrunning{F. Galati and M.A. Zuluaga}
%
\institute{Data Science Department\\
EURECOM, Sophia Antipolis, France\\
\email{\{galati,zuluaga\}@eurecom.fr}
}
\maketitle              
\begin{abstract}
Deep learning methods have reached state-of-the-art performance in cardiac image segmentation. Currently, the main bottleneck towards their effective translation into clinics requires assuring continuous high model performance and segmentation results. 
In this work, we present a novel learning framework to monitor the performance of heart segmentation models in the absence of ground truth. 
Formulated as an anomaly detection problem, the monitoring framework allows deriving surrogate quality measures for a segmentation and allows flagging suspicious results. We propose two different types of quality measures, a global score and a pixel-wise map. We demonstrate their use by reproducing the final rankings of a cardiac segmentation challenge in the absence of ground truth. Results show that our framework is accurate, fast, and scalable, confirming it is a viable option for quality control monitoring in clinical practice and large population studies.

\end{abstract}

\section{Introduction}
\label{sec:intro}
With the advent of learning-based techniques over the last decade, cardiac image segmentation has reached state-of-the-art performance \cite{Bernard2018}. 
This achievement has opened the possibility to develop image segmentation frameworks that can assist and automate (partially or fully) the image analysis pipelines of large-scale population studies or routine clinical procedures. 

The current bottleneck towards the large-scale use of learning-based pipelines in the clinic comes from the monitoring and maintenance of the deployed machine learning systems~\cite{Sculley2015}. As shown in \cite{Bernard2018}, despite the very high performances achieved, these methods may generate anatomically impossible results. In clinical practice and population studies, it is of utmost importance to constantly monitor a model's performance to determine when it degrades or fails, leading to poor quality results, as they may represent important risks. A system’s continuous performance assessment and the detection of its degradation are challenging after deployment, due to the lack of a reference or ground truth. Therefore, translation of models into clinical practice requires the development of monitoring mechanisms to measure a model's segmentation quality, in the absence of ground truth, which guarantee their safe use in clinical routine and studies. 

In a first attempt to assess performances of cardiac segmentation models in the absence of ground truth, 
Robinson et al.~\cite{Robinson2018} proposed a supervised DL-based approach to predict the Dice Score Coefficient (DSC). More recently, Puyol-Ant{\'{o}}n et al.~\cite{PuyolAnton2020} used a Bayesian neural network to measure a model's performance by classifying its resulting segmentation as correct or incorrect, whereas Ruijsink et al.~\cite{Ruijsink2020} also use qualitative labels to train a support vector machine that predicts both the quality of the segmentation and of derived functional parameters.
The main drawback of these methods is that they require annotations reflecting a large set of quantitative (e.g. DSC) or qualitative (e.g. correct/incorrect) segmentation quality levels, which can be difficult to obtain. The Reverse Classification Accuracy (RCA)~\cite{Robinson2019,Valindria2017} addresses this problem by using atlas label propagation. This registration-based method relies on the spatial overlap between the predicted segmentation and a reference atlas. It works under the hypothesis that 
if the predicted segmentation is of good quality, then it will produce a good segmentation on at least one atlas image. 
However, the atlas registration step may fail. This is often the case for certain cardiac pathologies that introduce significant morphological deformations that the registration step is not able to recover~\cite{Zuluaga2015b}. In such cases, it is necessary to verify the results and manually fine-tune the registration step, limiting the method's scalability.  

We present a novel learning framework to monitor the performance of cardiac image segmentation models in the absence of ground truth. The proposed framework is formulated as an anomaly detection problem. The intuition behind this work lies in the possibility of estimating a model of the variability of cardiac segmentation masks from a reference training dataset provided with reliable ground truth. This model is represented by a convolutional autoencoder, which can be subsequently used to identify anomalies in segmented unseen images. Differently from previous learning-based approaches~\cite{PuyolAnton2020,Robinson2018}, we avoid the need of any type of annotations about the quality of a segmentation for training. Our approach also avoids the required spatial alignment between ground truth images and segmentations of RCA~\cite{Robinson2019,Valindria2017}, thus circumventing image registration.



\section{Method}

\label{sec:format}

\begin{figure}[t]
	\centering
	\centerline{\includegraphics[width=\textwidth]{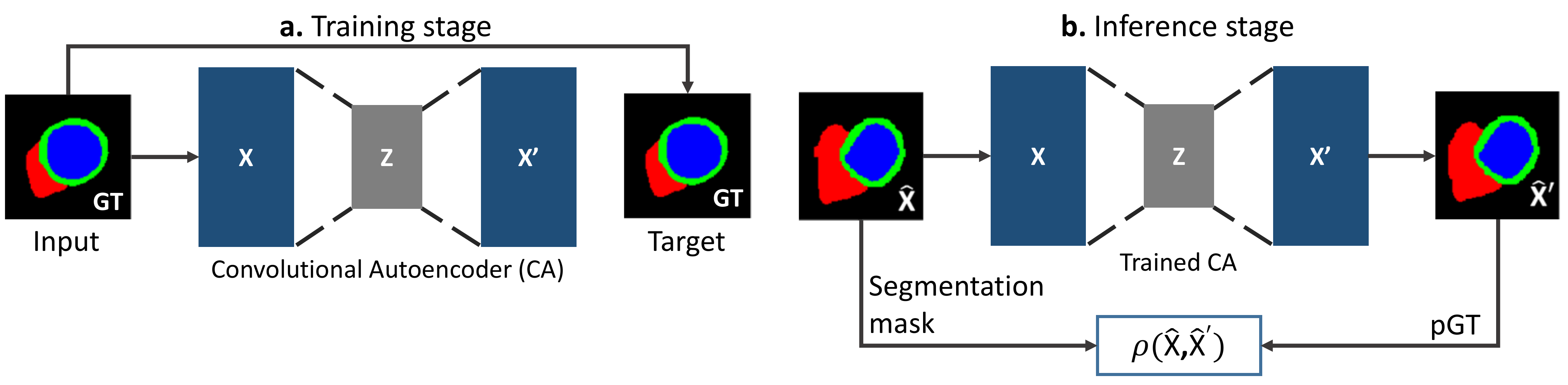}}
	\caption{\textbf{a.} A Convolutional Autoencoder (CA) is trained with ground truth (GT) masks from a cardiac imaging dataset. \textbf{b.} At inference, the CA reconstructs an input mask $\widehat{X}$, previously segmented by a model. The reconstructed mask $\widehat{X}'$ acts as pseudo ground truth (pGT) to estimate a function $\smash{\rho(\widehat{X},\widehat{X}')}$, a surrogate measure of the segmentation quality and the model's performance.}
	\label{fig:sketch}
\end{figure}

Let us denote $X \in \mathcal{C}^{H \times W}$ a segmentation mask of width $W$ and height $H$, with $\mathcal{C}$ the set of possible label values. A Convolutional Autoencoder (CA) is trained to learn a function $f: \mathcal{C}^{H \times W} \to \mathcal{C}^{H \times W}$, with $X'= f(X) \approx X$, by minimizing a global dissimilarity measure between an input mask $X$ and its reconstruction $X'$. In an anomaly detection setup, the CA is trained using normal samples, \textit{i.e.} samples without defects. In our framework, the normal samples are the ground truth (GT) masks associated with the images used to train a segmentation model (Fig.~\ref{fig:sketch}a). The CA learns to reconstruct defect-free samples, \textit{i.e.} the GT, through a bottleneck, the latent space $Z$.

At inference (Fig.~\ref{fig:sketch}b), the CA is used to obtain $\widehat{X}'= f(\widehat{X})$, where $\widehat{X}$ is a segmentation mask, generated by a cardiac segmentation model/method on unseen data, and $\widehat{X}'$ its reconstruction. Since the CA is trained with ground truth data, the quality of the reconstruction will be generally higher for segmentation masks with similar characteristics than those in the ground truth. Poor segmentations, which the CA has not encountered at training, will instead lead to bad reconstructions  ($\smash{\widehat{X}'\not\approx \widehat{X}}$). Autoencoder-based anomaly detection methods exploit the reconstruction error, i.e. $\smash{\|\widehat{X}'-\widehat{X}\|_2}$, to quantify how anomalous is a sample~\cite{Audibert2020}. We use this principle to establish a surrogate measure of the segmentation quality by quantifying a segmented mask and its reconstruction.     

Let us so formalize the function $\smash{\rho(\widehat{X},\widehat{X}')}$, a surrogate measure of the segmentation quality of the mask in the absence of GT. In this context, we denote $\smash{\widehat{X}'}$ a pseudo GT (pGT) since it acts as the reference to measure performance.
We present two different scenarios for $\rho$. First, we propose 
\begin{equation}\label{eq:typeone}
	\rho_1 : \mathcal{C}^{H \times W} \to \mathbb{R}, 
\end{equation}
which represents the most common setup in autoencoder-based anomaly detection. Due to the generic nature of $\rho_1$, well-suited metrics for segmentation quality assessment can be used, such as the DSC or the Hausdorff Distance (HD). Secondly, we propose 
\begin{equation}\label{eq:typetwo}
	\rho_2: \mathcal{C}^{H \times W} \to \mathbb{R}^{H \times W}.
\end{equation} 
This function generates a visual map of the inconsistencies between the two masks. We use as $\rho_2(\cdot)$ a pixel-wise XOR operation between the segmentation mask $\smash{\widehat{X}}$ and the pGT. 

These two types of measures can be used jointly for performance assessment and model monitoring. Measures obtained from $\rho_1$-type functions can be paired with a threshold to flag poor segmentation results. The raised alert would then be used to take application-specific countermeasures as, for instance, a visual inspection of an inconsistency map generated by $\rho_2$. 

\subsubsection{Network Architecture.}
\definecolor{conv}{HTML}{8d8794}
\definecolor{norm}{HTML}{8d8794}
\definecolor{relu}{HTML}{8d8794}
\definecolor{drop}{HTML}{8d8794}


\tikzset{conv/.style={black,draw=black,fill=conv!50,rectangle,minimum height=0.4cm,rounded corners}}
\tikzset{norm/.style={black,draw=black,fill=norm!50,rectangle,minimum height=0.4cm,rounded corners}}
\tikzset{relu/.style={black,draw=black,fill=relu!50,rectangle,minimum height=0.4cm,rounded corners}}
\tikzset{drop/.style={black,draw=black,fill=drop!50,rectangle,minimum height=0.4cm,rounded corners}}
\tikzset{back/.style={black,draw=black!50,fill=yellow!20,rectangle,minimum height=2.6cm,minimum width=3.0cm,dashed}}

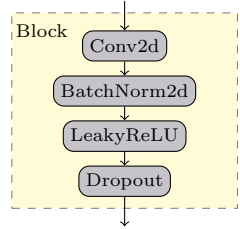
\begin{figure}
\begin{center}
\scriptsize
    \begin{minipage}[c]{.52\textwidth}
    \begin{tabularx}{0.92\linewidth}{lll |p{25pt}|p{25pt}|p{25pt}|}
        \toprule 
            Layer && Output Size & \multicolumn{3}{c}{Parameters}\\
        \cmidrule (lr){4-6} &&&
            \multicolumn{1}{c|}{Kernel} &  \multicolumn{1}{c|}{Stride}  & \multicolumn{1}{c|}{Padding}\\
        \midrule 
            Input && $256 \mathsf{x} 256 \mathsf{x} 4$ \\
            Block1 && $ 128 \mathsf{x} 128 \mathsf{x} 32$ & \,\quad$4 \mathsf{x} 4$ &\,\,\,\quad$2$&\,\,\,\quad$1$\\
            Block2 && $ 64 \mathsf{x} 64 \mathsf{x} 32$ &  \,\quad$4 \mathsf{x} 4$ &\,\,\,\quad$2$&\,\,\,\quad$1$\\
            Block3 && $32 \mathsf{x} 32 \mathsf{x} 32$ &  \,\quad$4 \mathsf{x} 4$ &\,\,\,\quad$2$&\,\,\,\quad$1$\\
            Block4 && $32 \mathsf{x} 32 \mathsf{x} 32$ &  \,\quad$3 \mathsf{x} 3$ &\,\,\,\quad$1$&\,\,\,\quad$1$\\
            Block5 && $16 \mathsf{x} 16 \mathsf{x} 64$ &  \,\quad$4 \mathsf{x} 4$ &\,\,\,\quad$2$&\,\,\,\quad$1$\\
            Block6 && $16 \mathsf{x} 16 \mathsf{x} 64$ &  \,\quad$3 \mathsf{x} 3$ &\,\,\,\quad$1$&\,\,\,\quad$1$\\
            Block7 && $8 \mathsf{x} 8 \mathsf{x} 128$ &  \,\quad$4 \mathsf{x} 4$ &\,\,\,\quad$2$&\,\,\,\quad$1$\\
            Block8 && $8 \mathsf{x} 8 \mathsf{x} 64$ &  \,\quad$3 \mathsf{x} 3$ &\,\,\,\quad$1$&\,\,\,\quad$1$\\
            Block9 && $8 \mathsf{x} 8 \mathsf{x} 32$ &  \,\quad$3 \mathsf{x} 3$ &\,\,\,\quad$1$&\,\,\,\quad$1$\\
            Conv2d && $4 \mathsf{x} 4 \mathsf{x} 100$ &  \,\quad$4 \mathsf{x} 4$ &\,\,\,\quad$2$&\,\,\,\quad$1$\\
        \bottomrule
    \end{tabularx}
    \end{minipage}
    \begin{minipage}[c]{.3\textwidth}
    \begin{tikzpicture}
        \pgfdeclarelayer{background}
        \pgfdeclarelayer{foreground}
        \pgfsetlayers{background,main,foreground}
        
        \node (s) at (0,-0.5) {};
        
        \node (block) at (-1.1,-1) {Block};
        \node[conv] (conv) at (0,-1.2) {Conv2d};
        \node[norm] (norm) at (0,-1.8) {BatchNorm2d};
        \node[relu] (relu) at (0,-2.4) {LeakyReLU};
        \node[drop] (drop) at (0,-3.0) {Dropout};
        
        \node (fs) at (0,-3.7) {};
        
        \draw[->] (s) -- (conv);
        \draw[->] (conv) -- (norm);
        \draw[->] (norm) -- (relu);
        \draw[->] (relu) -- (drop);
        \draw[->] (drop) -- (fs);
        
        \begin{pgfonlayer}{background}
            \path (relu.north)+(0,0.13) node[back] (back) {};
        \end{pgfonlayer}
    
    \end{tikzpicture}
    \end{minipage}
  
  \caption{Architecture of the encoding module. The decoder is built by reversing this structure and replacing convolutions with transposed convolutions. }\label{fig:ca}
\end{center}
\end{figure}

We use the CA architecture proposed in \cite{bergmann2019} as the backbone network (Fig.~\ref{fig:ca}) with the following modifications. We use a latent space dimension to accommodate 100 feature maps of size 4$\times$4. A softmax activation function is added to the last layer to normalize the output to a probability distribution over predicted output classes, as well as batch-normalization and dropout to each hidden layer. We use the loss function $\mathcal{L}=\mathcal{L}_{\text{MSE}}(X,X')+\mathcal{L}_{\text{GD}}(X,X')$, where $\mathcal{L}_{\text{MSE}}$ is the mean squared error loss and $\mathcal{L}_{\text{GD}}$ the generalized dice loss~\cite{sudre2017}. Trained over 500 epochs, for the first 10 epochs $\mathcal{L}_{\text{GD}}$ is computed leaving aside the background class to avoid the convergence to a dummy blank solution. The network weights are set using a \textit{He} normal initializer. The Adam optimizer is initialized with learning rate \num{2e-4} and a weight decay of \num{1e-5}. After every epoch, the model is evaluated on the validation set. The weights retrieving the lowest $\mathcal{L}$ value are stored for testing.

\section{Experiments and Results}
\label{sec:experiments}
Section~\ref{sec:setup} describes the datasets used, the setup and implementation of the experiments. Experimental results are then presented in section~\ref{sec:results}.
\subsection{Experimental Setup}\label{sec:setup}
\subsubsection*{Data.} We used data from the MICCAI 2017 Automatic Cardiac Diagnosis Challenge (ACDC)~\cite{Bernard2018}. The dataset consists of an annotated set with 100 short-axis cine magnetic resonance (MR) images, at end-diastole (ED) and end-systole (ES), with corresponding labels for the left ventricle (LV), right ventricle (RV), and myocardium (MYO). The set was split into training and validation subsets using an 80:20 ratio.  
The challenge also provides a testing set with 50 cases, with no ground truth publicly available. To have uniform image sizes, these were placed in the middle of a 256$\times$256 black square. Those exceeding this size were center cropped.

\subsubsection*{Setup.} We trained the monitoring framework using the ground truth masks from the ACDC training set and used it to assess the performance of five methods participating in the ACDC Challenge~\cite{baumgartner2017,isensee2017,khened2017,tziritas2017,yang2017} and an additional state-of-the-art cardiac segmentation model~\cite{bai2018}. We trained five models \cite{bai2018,baumgartner2017,isensee2017,khened2017,yang2017} using the challenge's full training set (MR images and masks) and then segmented the ACDC test images. For the remaining method~\cite{tziritas2017}, we obtained the segmentation masks directly from the participating team. 

The segmentations from every method were fed to the monitoring framework. The resulting pGTs were used to compute $\rho_1$-type measures (Def.~\ref{eq:typeone}), the DSC and the HD, and a $\rho_2$-type measure, an inconsistency map (Def.~\ref{eq:typetwo}). We also computed pseudo DSC/HD using the RCA~\cite{Valindria2017}. The ACDC challenge platform estimates different performance measures (DSC, HD, and other clinical measures) on the testing set upon submission of the segmentation results. We uploaded the masks from every model to obtain real DSC and HD. To differentiate the real measures computed by the platform from our estimates, we denote the latter ones pDSC and pHD. In our experiments, we set $\text{pHD} > 50$ or $\text{pDSC} < 0.5$ to flag a segmentation as suspicious and $\text{pHD} = \text{pDSC} = 0$ to raise an erroneous segmentation alert flag.

\subsubsection*{Implementation.} We implemented our framework in PyTorch. All the cardiac segmentation models used the available implementations, except for~\cite{tziritas2017} where we had the segmentation masks. The RCA was implemented following the guidelines in \cite{Valindria2017} using a previously validated atlas propagation heart segmentation framework~\cite{Zuluaga2013}. All experiments ran on Amazon Web Services with a Tesla T4 GPU. To encourage reproducibility, our code and experiments are publicly available from a Github repository\footnote{ \url{https://github.com/robustml-eurecom/quality\_control\_CMR}}.

\subsection{Results}\label{sec:results}
\begin{figure}[t]
	\centering
	\centerline{\includegraphics[width=\textwidth]{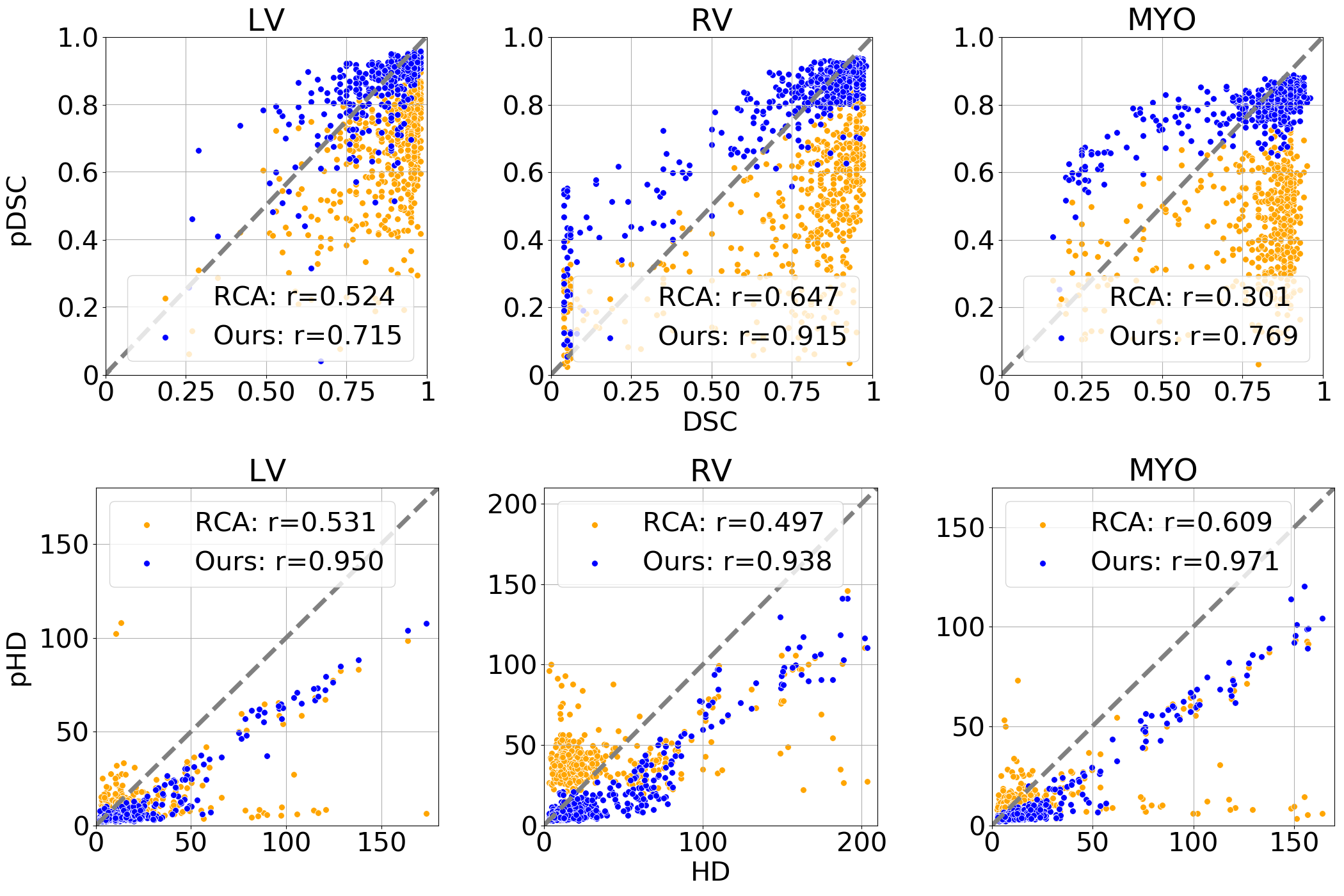}}
	\caption{DSC vs, pDSC (top) and HD vs. pHD (bottom) for our framework (blue dots) and RCA (yellow dots) on the left ventricle (LV), right ventricle (RV) and myocardium (MYO).}
	\label{fig:correlation}
\end{figure}

Figure~\ref{fig:correlation} presents scatter plots of the real DSC and HD from the ACDC platform and the pDSC and pHD obtained with our framework and the RCA~\cite{Robinson2019,Valindria2017}. We present results for LV, RV, and MYO over all generated segmentation masks, and report the Pearson correlation coefficient $r$. 

The results show a high positive correlation between real scores and our estimations. Our framework consistently outperforms RCA. For both RCA and our proposed approach, the real and pseudo HDs show stronger correlations than the DSC. This can be explained by the higher sensitivity of the HD to segmentation errors. Instead, the DSC shows little variability in the presence of minor segmentation, suggesting that both methods have difficulties in modeling small deviations from the reference ground truth data, i.e. very high-quality segmentations.  


Table~\ref{tab:comparison_acdc} simulates the ACDC Challenge results using real HD and pHD for every model to determine if our framework is a reliable means to rank the performance of the different cardiac segmentation methods. We do not use the pDSC for the ranking, as it reported a lower Pearson correlation coefficient $r$ in our first experiment (Fig.~\ref{fig:correlation}).

We report results for LV, RV, and MYO in ED and ES and compare them against the RCA. The ranking quality is assessed using Spearman's rank correlation coefficient $r_s$ between the real and the pseudo measures. The $r_s$ assesses if there is a monotonic relationship between both measures, i.e. it allows to determine if the pseudo measure is a valid criterion to rank the different methods. For fairness in the comparison, the ranking does not include one  method~\cite{isensee2017}, where RCA failed. In five out of six cases, we were able to perfectly reproduce the real ranking ($r_s$=1.0). In the remaining case, the left ventricle in end-diastole (ED), there is only one difference between the real and our pseudo ranking ($r_s$=0.9), where the 3rd and 4th places were swapped. The positive high  $r_s$ scores obtained by our framework confirm that it is a reliable mean for method ranking. 

\begin{table*}[t]
	\caption{ACDC Challenge simulation with 6 models using the real HD (GT) and the pHD obtained with our framework (Ours) and the RCA. HD and pHD reported in mm. The Spearman's rank correlation coefficient $r_s$ measures the ranking accuracy (the closer to 1.0 the better). pHD scores using RCA are excluded for \cite{isensee2017}, where the registration step failed.}
	\label{tab:comparison_acdc}
	\centering
	\begin{tabular}{l|r|r|r|r|r|r|r|r|r}
		\hline
		\multicolumn{10}{c}{\textbf{ED}}\\
		\hline
		& \multicolumn{3}{c|}{LV} &  \multicolumn{3}{c|}{RV} &  \multicolumn{3}{c}{MYO}\\
		\cline{2-10}
		Model & 
		\,\scalebox{.7}{GT} \,&
		\, \scalebox{.7}{\textbf{Ours}} \, &
		\, \scalebox{.7}{RCA} \, &
		\,\scalebox{.7}{GT} \,&
		\, \scalebox{.7}{Ours} \, &
		\, \scalebox{.7}{RCA} \, &
			\,\scalebox{.7}{GT} \,&
		\, \scalebox{.7}{\textbf{Ours}} \, &
		\, \scalebox{.7}{RCA} \, \\
		\hline
		Bai~\cite{bai2018} &
		39.01 & 23.38 & 15.55 &
		50.21 & 31.82 & 56.22 &
		47.10 & 28.46 & 20.42\\
		Baumgartner~\cite{baumgartner2017} &
		7.14 & 3.87 & 9.30 &
		14.00 & 7.72 & 37.63 &
		9.49 & 4.43 & 10.52\\
		Isensee~\cite{isensee2017} &
		7.01 & 3.88 & - &
		11.40 & 7.82 & - &
		8.44 & 4.38 & - \\
		Khened~\cite{khened2017} &
		16.81 & 6.39 & 10.58 &
		13.25 & 6.87 & 39.01 &
		16.09 & 6.08 & 11.22\\
		Tziritas~\cite{tziritas2017}&
		8.90 & 4.69 & 8.92 &
		21.02 & 9.86 & 41.10 &
		12.59 & 4.58 & 10.65\\
		Yang~\cite{yang2017} &
		16.95 & 5.29 & 12.96 &
		86.08 & 47.24 & 44.75 &
		31.93 & 16.39 & 15.12\\
		\hline
		$r_s$ & 
		- & 0.90 & 0.90 &
		- & 1.00 & 0.80 &
		- & 1.00 & 1.00\\
		\hline
		\hline
		\multicolumn{10}{c}{\textbf{ES}}\\
		\hline
		& \multicolumn{3}{c|}{LV} &  \multicolumn{3}{c|}{RV} &  \multicolumn{3}{c}{MYO}\\
		\cline{2-10}
		Model & 
		\,\scalebox{.7}{GT} \,&
		\, \scalebox{.7}{\textbf{Ours}} \, &
		\, \scalebox{.7}{RCA} \, &
		\,\scalebox{.7}{GT} \,&
		\, \scalebox{.7}{\textbf{Ours}} \, &
		\, \scalebox{.7}{RCA} \, &
		\,\scalebox{.7}{GT} \,&
		\, \scalebox{.7}{\textbf{Ours}} \, &
		\, \scalebox{.7}{RCA} \, \\
		\hline
		Bai~\cite{bai2018} &
		50.53 & 29.56 & 20.01 &
		52.73 & 31.40 & 53.68 &
		52.72 & 31.05 & 26.60\\
		Baumgartner~\cite{baumgartner2017} &
		10.51 & 4.41 & 9.56 &
		16.32 & 7.10 & 35.50 &
		12.47 & 4.77 & 9.33\\
		Isensee~\cite{isensee2017} &
		7.97 & 4.07 & - &
		12.07 & 6.99 & - &
		7.95 & 4.27 & - \\
		Khened~\cite{khened2017} &
		20.14 & 6.96 & 11.72 &
		14.71 & 7.07 & 35.65 &
		16.77 & 6.03 & 10.36\\
		Tziritas~\cite{tziritas2017}&
		11.57 & 5.00 & 10.46 &
		25.70 & 9.61 & 36.51 &
		14.78 & 5.59 & 10.60\\
		Yang~\cite{yang2017} &
		19.13 & 6.11 & 11.78 &
		80.42 & 33.21 & 40.68 &
		32.54 & 16.98 & 13.68\\
		\hline
		$r_s$&- & 1.00 & 0.90 &
		- & 1.00 & 0.80 &
		- & 1.00 & 0.90\\
		\hline
	\end{tabular}

\end{table*}

\begin{figure}[!t]
	\centering
	\includegraphics[width=\textwidth]{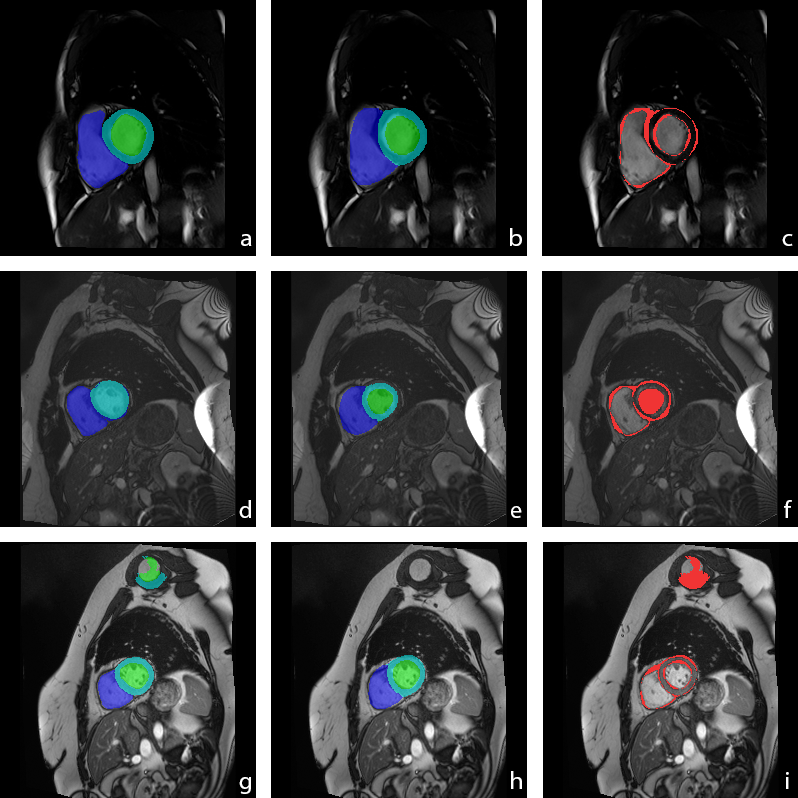}
	\caption{Segmentation masks (a, d and g), pGTs (b, e and h), and inconsistency maps (c, f and i) for \textbf{first row:} successful segmentation, according to the alert thresholds; \textbf{middle row:} a segmentation mask flagged as erroneous with $\text{pHD} = \text{pDSC} = 0$ in the LV; and \textbf{bottom row:} a segmentation mask flagged as suspicious with $\text{pHD} = 104.93$ for the LV ($\text{pDSC} = 0.814$). The inconsistency maps confirm the errors.}
	\label{fig:examples}
\end{figure}

Through the use of alert flags we were able to detect 16 cases for which the challenge platform had reported \verb|NaN| values indicating errors in the submitted results. Fifteen cases were flagged as erroneous ($\text{pHD} = \text{pDSC} = 0$) and one as suspicious ($\text{pHD} > 50$) by our framework. Fig.~\ref{fig:examples} middle and bottom row illustrates two of these cases. The middle row presents a segmentation flagged as erroneous, where the inconsistency map confirms that the left ventricle has not been segmented. The bottom row shows the case of a segmentation flagged as suspicious, where the left ventricle's pHD is high ($\text{pHD} = 104.93$), although the $\text{pDSC} = 0.814$ is within normal range. The inconsistency map confirms the clear segmentation error. 

Additionally, Fig.~\ref{fig:examples} top row illustrates an example of a segmentation mask that has not been flagged, i.e. a segmentation result considered good by the $\rho_1$-type metrics of our framework ($\text{pHD} < 50$ and $\text{pDSC} > 0.5$ for all LV, RV and MYO). The inconsistency map (Fig.~\ref{fig:examples}c) flags pixels in the edges of all the structures as suspicious. While it is clear from the images that the segmentation mask for the RV ($\text{pDSC}=0.75$, $\text{DSC}=0.89$) has some errors in the edges, which require a revision, the flagged pixels in the LV and MYO are more difficult to assess. In such a case, a more informative $\rho_2$-type measure than the selected XOR is desirable.


\section{Discussion and Conclusions}\label{sec:conclusion}
We presented a novel learning framework to monitor the performance of cardiac image segmentation models in the absence of ground truth. Our framework addresses the limitations of previous learning-based approaches~\cite{Kohlberger2012,PuyolAnton2020,Robinson2018,Ruijsink2020} thanks to its formulation under an anomaly detection paradigm that allows training without requiring quality scores labels. Our results show a good correlation between real performance measures and those estimated with the pseudo ground truth (pGT), making it a reliable alternative when there is no reference to assess a model. Compared with state-of-the-art RCA~\cite{Robinson2019,Valindria2017}, our method avoids the use of image registration which makes it more robust, scalable, and considerably faster ($\sim20$ min RCA vs. $\sim 0.2$ s ours, per case). CAs allow for fast inference which conforms to real-time use, thus permitting a quick quality assignment, for example, in a clinical setting. All these characteristics make the proposed framework a viable option for quality control and system monitoring in clinical setups and large population studies.

A current limitation of the proposed framework is that it is less reliable when assessing high-quality segmentations. A simple and practical way to address this could be to increase the alert thresholds (lower pHDs, higher pDSCs), making them more sensitive to small segmentation errors, and to develop more informative $\rho_2$-type measures, similar to those proposed by uncertainty quantification methods~\cite{PuyolAnton2020}. However, we consider that a more principled approach should be favored by investigating mechanisms to model these smaller deviations from the reference.  

Another straightforward extension of this work could be to embed the proposed framework within the segmenter network, as a way to perform quality control during training. This idea has been previously explored by different AE-based segmentation  methods~\cite{oktay2017,painchaud2020,yue2019}. However, the results reported in~\cite{painchaud2020} suggest that this does not fully solve the problem of unexpected erroneous and anatomically impossible results, originally reported in~\cite{Bernard2018}. Therefore, we consider that separate and modular monitoring frameworks should be favored over end-to-end solutions and future works should focus on improving the limitations of the current quality control techniques, in particular, their poor sensitivity to smaller errors. 

\subsubsection*{Acknowledgments.} The authors would like to thank  Christian Baumgartner, Elios Grinias, Jelmer M. Wolterink, and Clement Zotti for their help in the reproduction of the ACDC Challenge rankings by sharing their code or results and, overall, through their valuable advice.

\bibliographystyle{splncs04}
%
\bibliography{main}
\end{document}